\pdfoutput=1

\documentclass[11pt]{article}
\usepackage{amsmath}
\usepackage[noend]{algpseudocode}
\usepackage{algorithm}
\usepackage{times}
\usepackage{amssymb}
\usepackage{booktabs} 
\usepackage{siunitx} 
\usepackage[table]{xcolor}
\usepackage{ACL2023}
\usepackage{times}
\usepackage{latexsym}
\usepackage{hyperref}
\usepackage{graphicx}
\usepackage[T1]{fontenc}

\usepackage[utf8]{inputenc}

\usepackage{microtype}

\usepackage{inconsolata}

\usepackage{subcaption}

\title{Speaking in Wavelet Domain: A Simple and Efficient Approach to Speed up Speech Diffusion Model}

\author{Xiangyu Zhang$^{1*}$, Daijiao Liu$^{1*}$, Hexin Liu$^{3}$, Qiquan Zhang$^{1}$\\ \textbf{Hanyu Meng}$^{1}$, \textbf{Leibny Paola Garcia}$^{4}$, \textbf{Eng Siong Chng}$^{3}$, \textbf{Lina Yao}$^{2}$\\
The University of New South Wales$^{1}$,
Data61 CSIRO$^{2}$ \\
Nanyang Technological University$^{3}$,
HLTCOE and Johns Hopkins University$^{4}$
}

\newcommand\nnfootnote[1]{%
  \begin{NoHyper}
  \renewcommand\thefootnote{}\footnote{#1}%
  \addtocounter{footnote}{-1}%
  \end{NoHyper}
}

\begin{document}
\maketitle

\begingroup
\def\thefootnote{$\star$}
\nnfootnote{*Equal contribution}
\endgroup

\begin{abstract}
Recently, Denoising Diffusion Probabilistic Models (DDPMs) have attained leading performances across a diverse range of generative tasks. However, in the field of speech synthesis, although DDPMs exhibit impressive performance, their long training duration and substantial inference costs hinder practical deployment. Existing approaches primarily focus on enhancing inference speed, while approaches to accelerate training—a key factor in the costs associated with adding or customizing voices—often necessitate complex modifications to the model, compromising their universal applicability. To address the aforementioned challenges, we propose an inquiry: \textbf{is it possible to enhance the training/inference speed and performance of DDPMs by modifying the speech signal itself?} In this paper, we \textbf{double} the training and inference speed of Speech DDPMs by simply redirecting the generative target to the wavelet domain. This method not only achieves comparable or superior performance to the original model in speech synthesis tasks but also demonstrates its versatility. By investigating and utilizing different wavelet bases, our approach proves effective not just in speech synthesis, but also in speech enhancement.
\end{abstract}

\section{Introduction}
Recently, with the advancement of deep learning, generative models have made significant progress in various fields~\cite{karras2019style,oord2016wavenet,yang2019pointflow}. Particularly, the emergence of diffusion models has elevated the capabilities of deep generative models to a new level~\cite{ho2020denoising,song2020score}. In the field of speech processing, Denoising Diffusion Probabilistic Models (DDPMs) not only exhibit astonishing performance in speech synthesis~\cite{kong2020diffwave,jeong2021diff} but also demonstrate commendable results in speech enhancement~\cite{lu2022conditional,yen2023cold}. However, despite the impressive results achieved by DDPMs in the field of speech processing, the requirement to generate a guarantee of high sample quality — typically necessitating hundreds to thousands of denoising steps — results in training and inference speeds that are daunting in practical applications.

\begin{figure}[t]
    \centering
    \includegraphics[width=0.45\textwidth]{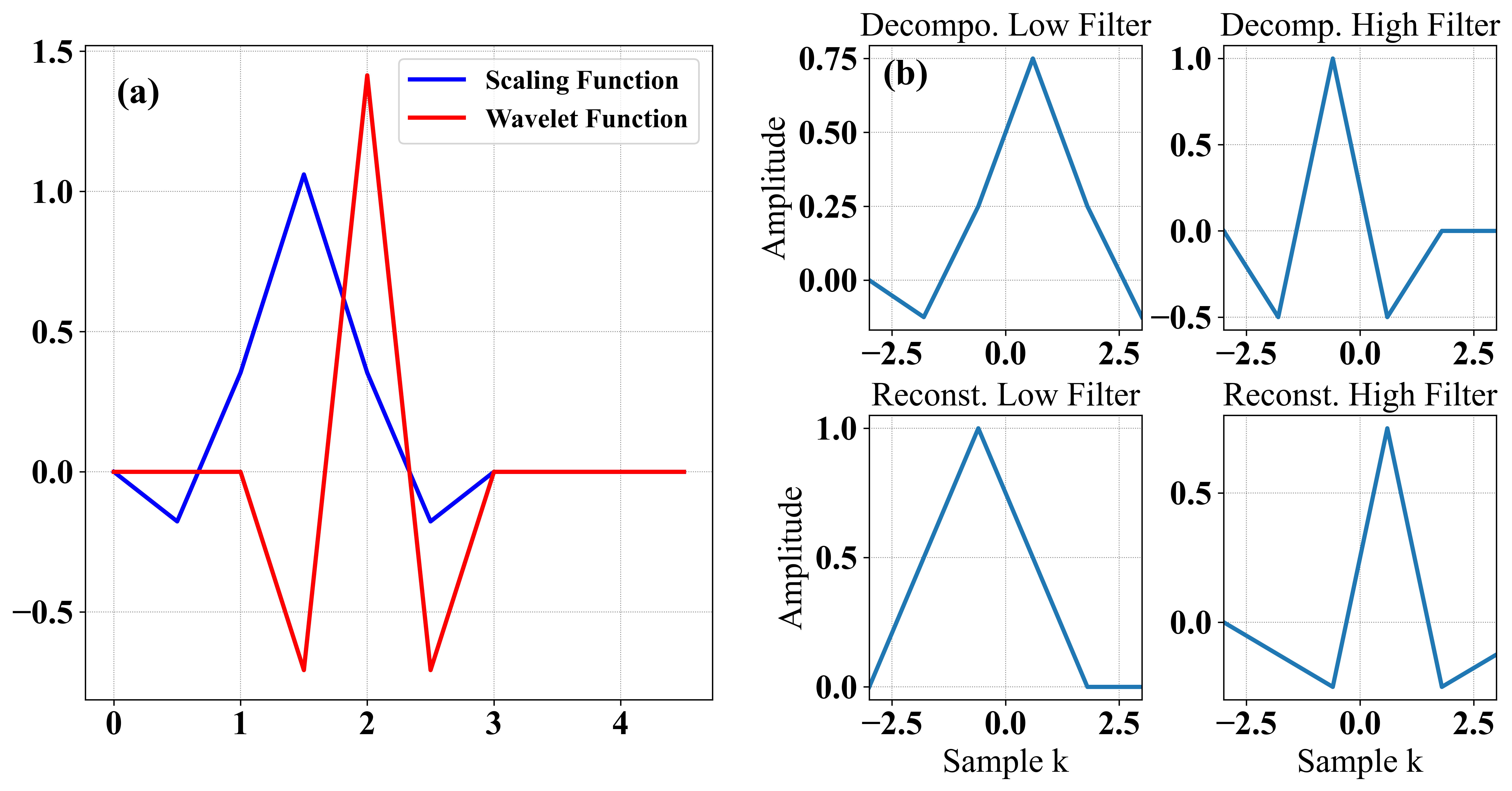}
    \caption{Wavelet of Cohen-Daubechies-Feauveau 5-tap/3-tap. (a) Scaling and wavelet functions, (b) decomposition and reconstruction filters.}\vspace{-6mm}
    \label{fig:wavelet1}
\end{figure}

Given these issues, researchers from various fields have attempted different methods to improve diffusion models. In the realm of speech processing, existing approaches have endeavored to alter the model structure to accelerate the inference speed of speech synthesis~\cite{huang2022fastdiff}, while others have experimented with changing training strategies to reduce the number of inference steps required for diffusion models in speech enhancement~\cite{lay2023single}. These approaches primarily focus on enhancing the inference speed of speech diffusion models. However, in the field of speech synthesis, the industry frequently requires incorporating new voices to accommodate varied requirements. Additionally, generative-based speech enhancement often demands tailoring models to distinct scenarios, which introduces practical limitations to the aforementioned methods in real-world applications.
\begin{figure*}[ht]
    \centering
    \includegraphics[width=1\textwidth]{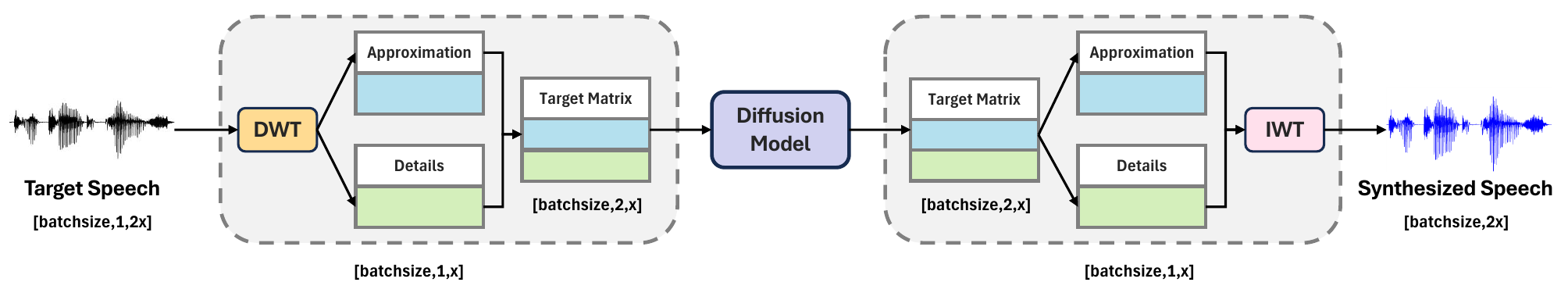}
    \caption{Overview of the Speech Wavelet Diffusion Model pipeline: First, the speech signal is decomposed into Approximation coefficients Matrix(cA) and Detail coefficients matrix(cD), the Diffusion model subsequently generates cA and cD and restores the speech signal from these matrices.} 
    \label{fig:overview}
\end{figure*}
In the field of computer vision, researchers have attempted to accelerate diffusion models using wavelets. Their efforts are mainly concentrated on score-based diffusion models~\cite{song2020score,song2021maximum}, employing wavelets to modify the training strategy, thereby simultaneously enhancing both training and inference speeds~\cite{guth2022wavelet}. \textbf{However, there is a significant difference between audio and image signals.} Unlike the common feature sizes of 64x64 or 256x256 in images, speech signals often have large feature sizes to ensure training quality. This means that the challenges in training speech models often stem from the nature of the speech signal itself~\cite{radford2023robust}. Considering this, we propose a question from a different angle: can we improve the training and inference speeds of DDPMs and significantly alleviate GPU memory pressure by operating directly on the speech signal itself?

The principle of simplicity often underlies effective methods, as evidenced by tools like LoRA~\cite{hu2021lora} and Word2Vec~\cite{mikolov2013efficient}. Inspired by the successful application of latent space diffusion models~\cite{rombach2022high} and wavelets in image compression~\cite{taubman2002jpeg2000}, we pivot the generative aim of speech DDPMs towards the compressed speech signal in the wavelet domain. This involves decomposing the speech signal using the Discrete Wavelet Transform(DWT) into high-frequency and low-frequency components. These components are then concatenated to form a unified generative target for our model. Through this approach, the feature-length of the data is halved, which enhances the GPU's parallel processing capabilities and significantly reduces the demand for GPU memory.

In the Further Study chapter, we have developed two additional modules: the Low Frequency Enhancer and the Multi-Level Accelerator. The former enhances low-frequency signals, allowing our method to not only double the speed compared to the original model but also achieve better performance. The latter, by integrating the Low-Frequency Enhancer with multi-level wavelet transform, further compress the speech signal. This enables an acceleration of more than five times while maintaining comparable results.

In summary, our contributions include the following:
\begin{itemize}
\vspace{-3mm}
\item We designed a simple, effective, and universal method that \textbf{doubles the training and inference speed} of the original model \textbf{without altering its architecture} while maintaining comparable performance. Testing across different models and tasks not only confirmed the wide applicability and versatility of our approach but also demonstrated that the Diffusion Models can generate speech components in the wavelet domain.
\vspace{-3mm}
\item  We designed two simple and easily integrable front-end modules. The first achieves \textbf{better performance than the original model while doubling the speed}. The second offers a performance comparable to the original while enabling an acceleration of more than five times.
\vspace{-8mm}
\item We offer a new perspective on accelerating and optimizing speech models by focusing on processing the signal itself rather than modifying the model, thereby charting a new course for future research.
\end{itemize}

\vspace{-0.4cm}
\section{Related Work}
\vspace{-0.1cm}
\begin{figure*}[ht]
    \begin{subfigure}[b]{0.3\textwidth}
        \centering
        \includegraphics[width=\linewidth]{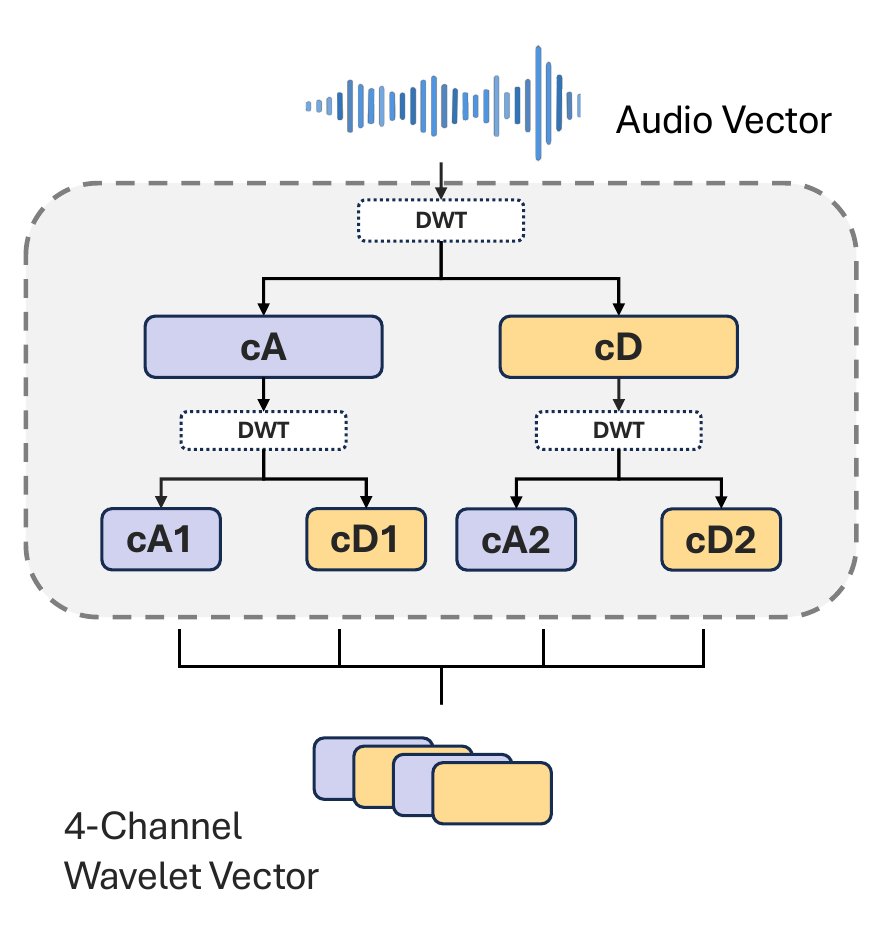}
        \caption{Block of Multi-Level Discrete Wavelet Transform}
        \label{fig:dwt}
    \end{subfigure}
    \hfill 
    \begin{subfigure}[b]{0.3\textwidth}
        \centering
        \includegraphics[width=\linewidth]{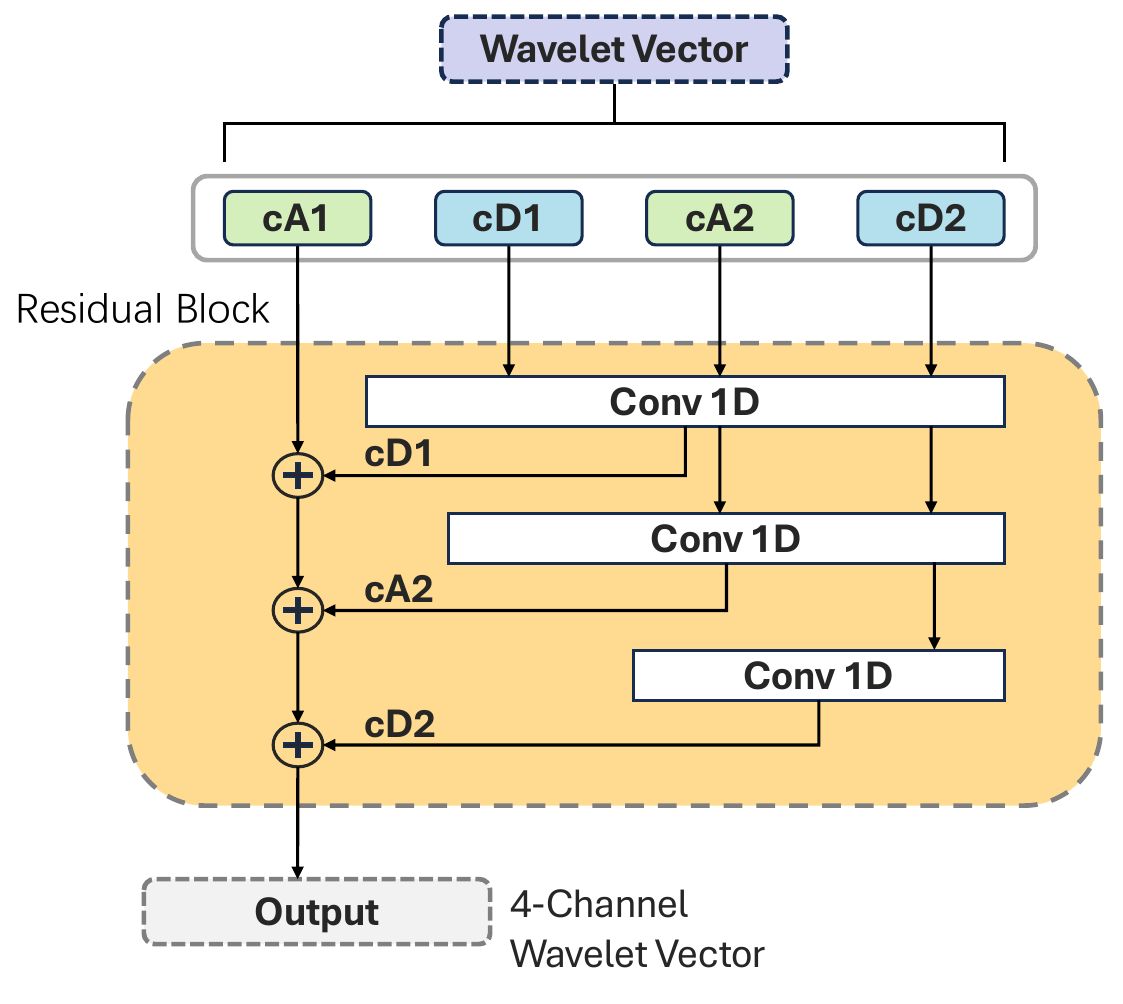}
        \caption{Multi-Level Low-Frequency Voice Enhancement Module}
        \label{fig:mulcnn}
    \end{subfigure}
    \hfill 
    \begin{subfigure}[b]{0.3\textwidth}
        \centering
        \includegraphics[width=\linewidth]{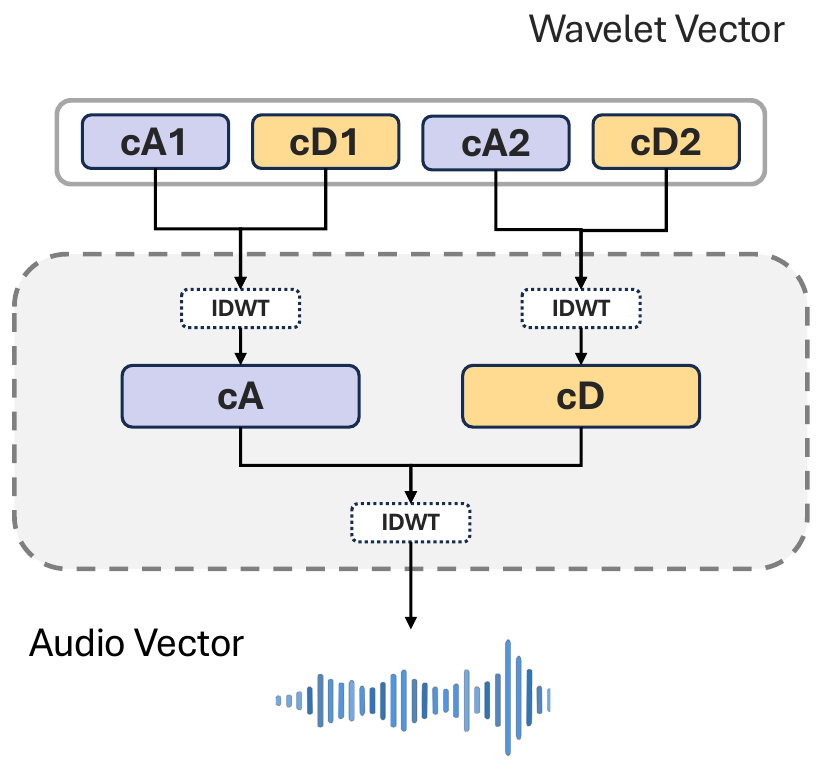}
        \caption{Block of Multi-Level Inverse Discrete Wavelet Transform}
        \label{fig:iwt}
    \end{subfigure}
    \caption{Overview of (a) Block of Multi-Level Discrete Wavelet Transform, (b) Multi-Level Low-Frequency Voice Enhancement Module, (c) Block of Multi-Level Inverse Discrete Wavelet Transform.}
    \vspace{-6mm}
    \label{further study overview}
\end{figure*}

\textbf{Diffusion Probabilistic Models}. Diffusion probabilistic models (DMs)~\cite{sohl2015deep,ho2020denoising} are a powerful and effective class of generative models, which are highly competitive in terms of sample quality, surpassing Variational Autoencoders (VAEs) and Generative Adversarial Networks (GANs) to become the state-of-the-art in a variety of synthesis tasks~\cite{dhariwal2021diffusion,liu2022diffsinger}. DMs comprise a forward noise diffusion process and a Markovian reverse diffusion process. They function by training a deep neural network to denoise content that has been corrupted with various levels of Gaussian noise. In the sampling phase, a generative Markov chain process based on Langevin dynamics~\cite{song2019generative} iteratively denoises from complete Gaussian noise to progressively generate the target samples. Due to their iterative nature, DMs experience a significant increase in training and sampling time when generating high-dimensional data~\cite{song2020denoising}. 
\newline
\textbf{Speech Synthesis}. In recent times, a variety of neural text-to-speech (TTS) systems have been developed~\cite{oord2016wavenet,binkowski2019high,valle2020flowtron,chen2024enhancing}. Initially, these systems generate intermediate representations, such as mel spectrograms or hidden representations, conditioned on textual input. This is followed by the use of a neural vocoder for the synthesis of the raw audio waveform. The pivotal role in the recent advancements of speech synthesis has been played by neural vocoders. Models like WaveFlow~\cite{ping2020waveflow} and WaveGlow~\cite{prenger2019waveglow} achieve training through likelihood maximization. On the other hand, models based on VAEs and GANs diverge from likelihood-centric models, often necessitating additional training losses to enhance audio fidelity. Another notable approach is the diffusion-based model~\cite{kong2020diffwave}, which stands out by synthesizing high-quality speech using a singular objective function. Our experiment will be conducted on a diffusion-based vocoder.
\newline    
\textbf{Speech Enhancement}. Speech enhancement is a field in audio signal processing focused on improving the quality of speech signals in the presence of noise~\cite{benesty2006speech}.  Recent advances in deep learning have significantly improved the performance of speech enhancement systems, enabling more effective noise suppression and clarity in diverse environments~\cite{zhang2020deepmmse,sun2023lightweight,zhang2024mamba}. In the realm of speech denoising, diffusion-based models are being effectively utilized. Lu~\cite{lu2022conditional} investigates the efficacy of diffusion model with noisy mel band inputs for this purpose. In a similar vein, Joan~\cite{serra2022universal} examines the application of score-based diffusion models for enhancing speech quality. Furthermore, Welker~\cite{welker2022speech} proposes formulations of the diffusion process specifically designed to adapt to real audio noises, which often present non-Gaussian properties.
\newline
\textbf{Speed Up Generative Speech Model}. Numerous efforts have been made to expedite speech synthesis, with Fastspeech~\cite{ren2019fastspeech} and Fastspeech 2~\cite{ren2020fastspeech} being among the most notable, both accelerating the process using transformer models. FastDiff~\cite{huang2022fastdiff}, a more recent development, aims to address the slow inference speed of diffusion models in practical applications, focusing primarily on hastening inference time. In contrast, our technology is designed \textbf{not only to accelerate both training and inference but also to be easily adaptable to various speech synthesis models}.
\section{Methodology}
\vspace{-0.3cm}
In this section, the proposed method is illustrated using the Cohen-Daubechies-Feauveau 5/3 wavelet as a case study~\cite{le1988sub}. We first explain how we utilize wavelet transforms for compressing and parallel processing of speech signals. Then, we delve into the specifics of accelerating speech synthesis and enhancement tasks.
\begin{algorithm}[t]
\begin{algorithmic}
\caption{Wavelet Diffwave Training}
\For{$i = 1,2,\dots,N_{\text{iter}}$}
    \State Sample $x_0 \sim q_{\text{data}}, \epsilon \sim \mathcal{N}(0, \mathbf{I}),$ and
    \State $t \sim \text{Uniform}(\{1,\dots,T\})$
    \State $y_0 = DWT(x_0)$
    \State Take gradient step on
    \State $\nabla_{\theta} \|\epsilon - \epsilon_{\theta}(\sqrt{\bar{\alpha}_t}y_0 + \sqrt{1-\bar{\alpha}_t}\epsilon, t)\|_2^2$
\EndFor
\label{diffwave training}
\end{algorithmic}
\end{algorithm}
\vspace{-0.2cm}
\subsection{Wavelet Transform and Compression}
\vspace{-0.1cm}
\label{WaveletCompression}
The Wavelet Transform is a key method in image compression, involving Discrete Wavelet Transform (DWT) and Inverse Discrete Wavelet Transform (IWT) to separate low-frequency (cA) and high-frequency (cD) components from signals~\cite{sullivan2003general}. We focus on the Daubechies-Feauveau 5/3 wavelet, shown in Figure~\ref{fig:wavelet1}, a biorthogonal wavelet commonly used in lossless compression algorithms~\cite{taubman2002jpeg2000}.
Let us define $L = \left[-\frac{1}{8}, \frac{2}{8}, \frac{6}{8}, \frac{2}{8}, -\frac{1}{8}\right]$ and $H = \left[\frac{1}{2}, 1, \frac{1}{2}\right]$ as the low-pass and high-pass filters, respectively. In the DWT Process, these filters are employed to decompose speech signals $x \in \mathbb{R}^{1\times 2x}$ into matrices $cA \in \mathbb{R}^{1 \times x}$ and $cD \in \mathbb{R}^{1 \times x}$. Subsequently, these matrices are concatenated to form $y \in \mathbb{R}^{2 \times x}$, as depicted in the left part of Figure~\ref{fig:overview}. In the IWT process, the matrix $y \in \mathbb{R}^{2 \times x}$ is divided back into $cA \in \mathbb{R}^{1 \times x}$ and $cD \in \mathbb{R}^{1 \times x}$, which are then reconstructed into the speech signal. The details of how Wavelet compresses speech and accelerates the model can be seen in Appendix~\ref{Wavelet Diffusion Accelerator}.\vspace{-0.2mm}
\vspace{-0.2cm}
\subsection{Wavelet-based Speech Diffusion Scheme}
\vspace{-0.1cm}
\subsubsection{Speech Synthesis}
\vspace{-0.1cm}
We evaluated our method using Diffwave~\cite{kong2020diffwave}, a well-known diffusion vocoder widely adopted in numerous TTS systems. 
We altered only the first layer of the one-dimensional convolutional network used for processing the input signal, ensuring that the number of channels remains constant, thereby keeping the network width unchanged in comparison with Diffwave. During the training process, the diffusion process is characterized by a fixed Markov chain transitioning from the concatenated wavelet data $y_0$ to the latent variable $y_T$. This is achieved via
\begin{equation}
\setlength{\abovedisplayskip}{4pt}
\setlength{\belowdisplayskip}{4pt}
\textstyle
    q(y_1, \ldots, y_T | y_0) = \prod_{t=1}^{T} q(y_t | y_{t-1}),
\end{equation}
where $q(y_t | y_{t-1})$ is defined as a Gaussian distribution $\mathcal{N}(y_t; \sqrt{1 - \beta_t}y_{t-1}, \beta_t \mathbf{I})$ and $\beta$ is a small positive constant. The function $q(y_t | y_{t-1})$ introduces slight Gaussian noise into the distribution of $y_{t-1}$, effectively adding minimal Gaussian noise to both $cA$ and $cD$. 
\begin{algorithm}[t]
\begin{algorithmic}
\caption{Wavelet Diffwave Sampling}
\State Sample $ye_T \sim p_{\text{latent}} = \mathcal{N}(0, \mathbf{I})$
\For{$t = T, T-1, \dots, 1$}
    \State Compute $\mu_{\theta}(y_t, t)$ and $\sigma_{\theta}(y_t, t)$ 
    \State Sample $y_{t-1} \sim p_{\theta}(y_{t-1}|y_t) = $
    \State $\quad \mathcal{N}(y_{t-1}; \mu_{\theta}(y_t, t), \sigma_{\theta}(y_t, t)^2\mathbf{I})$
\EndFor
\State $x_0 = IWT(y_0)$
\State \textbf{return} $x_0$
\label{diffwave sampling}
\end{algorithmic}
\end{algorithm}

The reverse process is characterized by a Markov chain transitioning from $y_T$ back to $y_0$. This is parameterized by $\theta$ and computed via
\begin{equation}
\setlength{\abovedisplayskip}{4pt}
\setlength{\belowdisplayskip}{4pt}
\textstyle
    p_{\theta}(y_0, \ldots, y_{T-1} | y_T) = \prod_{t=1}^{T} p_{\theta}(y_{t-1} | y_{t}).
\end{equation}

The distribution $p(y_T)$ originates from an isotropic Gaussian and is composed of two distinct components, corresponding respectively to $cA$ and $cD$. The term $p_{\theta}(y_{t-1} | y_{t})$ is parameterized by a Gaussian distribution $\mathcal{N}(y_{t-1}; \mu_{\theta}(y_t, t), \sigma_{\theta}(y_t, t)^2\mathbf{I})$. Here, $\mu_{\theta}$ yields a $2 \times X$ matrix representing the mean values for $cA$ and $cD$, while $\sigma_{\theta}$ produces two real numbers, indicating the standard deviations for $cA$ and $cD$.

The training objective is to minimize the following unweighted variant of the variational lower bound (ELBO):
\begin{equation}
\setlength{\abovedisplayskip}{4pt}
\setlength{\belowdisplayskip}{4pt}
\scalebox{0.85}{$
    \min_{\theta} L(\theta) = \mathbb{E} \left\| \epsilon - \theta\left(\sqrt{\overline{\alpha}_t} y_0 
    + \sqrt{1 - \overline{\alpha}_t} \epsilon, t\right) \right\|^2
$}
\end{equation}

where $\overline{\alpha}_t$ is derived from the variance schedule, parameter $\theta$ denotes a neural network that outputs noise for both $cA$ and $cD$. Furthermore, $\epsilon$ is represented as a $2 \times X$ matrix, encapsulating the actual noise values corresponding to both $cA$ and $cD$. The detailed procedures for training and sampling are outlined in Algorithm~\ref{diffwave training} and Algorithm~\ref{diffwave sampling}.

\vspace{-0.2mm}
\subsubsection{Speech Enhancement}
\vspace{-0.1mm}
We also evaluated our algorithm in Diffusion-based Speech Enhancement tasks, employing CDiffuSE~\cite{lu2022conditional} as a test case to demonstrate the effectiveness of our approach. Their diffusion forward process after wavelet processing can be formulated as
\begin{equation}
\setlength{\abovedisplayskip}{4pt}
\setlength{\belowdisplayskip}{4pt}
\begin{split}
   q_{\text{diff}}(y_t | y_0, y_{n}) & = \mathcal{N}\left (y_t; (1 - m_t) \sqrt{\bar{\alpha}_t} y_0 + \right. \\
   & \left.  m_t \sqrt{\bar{\alpha}_t} y_{n}, \delta_t \mathbf{I} \right ).
\end{split}
\end{equation}
The variable $m_t$ represents the interpolation ratio between the clean wavelet data $y_0$ and the noisy wavelet data $y_{n}$. This ratio initiates at $m_0 = 0$ and progressively increases to $m_t = 1$. The term $\bar{\alpha}_t$ is computed following the same methodology as employed in Diffwave, and $\delta_t$ is defined as $(1 - \alpha_t) - m_t^2\alpha_t$. The reverse process is formulated as
\begin{equation}
\setlength{\abovedisplayskip}{4pt}
\setlength{\belowdisplayskip}{4pt}
\scalebox{0.85}{$
    p_{\theta}(y_{t-1} | y_t, y_{n}) = \mathcal{N}(y_{t-1}; \mu_{\theta}(y_t, y_{n}, t), \tilde{\delta}_t \mathbf{I}),
$}
\end{equation}
where $\mu_{\theta}(y_t, y_{noise}, t)$ is the mean of a linear combination of $y_t$ and $y_{noise}$, being formulated as
\begin{equation}
\setlength{\abovedisplayskip}{4pt}
\setlength{\belowdisplayskip}{4pt}
\scalebox{0.85}{$
    \mu_{\theta}(y_t, y_n, t) = c_{y_t} y_t + c_{y_n} y_n - c_{\epsilon_t} \epsilon_{\theta}(y_t, y_n, t).
$}
\end{equation}

Parameters $c_{y_t}$, $c_{y_n}$, and $c_{\epsilon_t}$ are derived from the ELBO optimization. The detailed procedures for training and sampling are outlined in Algorithm~\ref{Wavelet CDiffuSE Training} and Algorithm~\ref{Wavelet CDiffuSE Sampling}. The details of coefficients and ELBO optimization can be seen in Appendix~\ref{Appendix CDiffuSE}.
\begin{algorithm}[t]
\caption{Wavelet CDiffuSE Sampling}
\begin{algorithmic}[1]
\State Sample $y_T \sim \mathcal{N}(y_T, \sqrt{\bar{\alpha}_T} y_n, \delta_T \mathbf{I})$
\For{$t = T, T - 1, \dots, 1$}
    \State Compute $c_{x_t}, c_{y_t}$ and $c_{\epsilon_t}$
    \State Sample $y_{t-1} \sim p_{\theta}(y_{t-1}|y_t, y_n) = \mathcal{N}(y_{t-1}; c_{x_t} y_t + c_{y_t} y_n - c_{\epsilon_t} \epsilon_{\theta}(y_t, y_n, t), \delta_t \mathbf{I})$
\EndFor
$x_0 = IWT(y_0)$
\State \textbf{return} $x_0$
\end{algorithmic}
\label{Wavelet CDiffuSE Sampling}
\end{algorithm}
\vspace{-0.2cm}
\section{Experiments}
\vspace{-0.2cm}
\subsection{Dataset}
\vspace{-0.1cm}
\textbf{Speech Synthesis} Our experiments were conducted using the LJSpeech dataset~\cite{ljspeech17}, comprising 13,100 English audio clips along with their corresponding text transcripts. The total duration of the audio in this dataset is approximately 24 hours. For the purpose of objectively assessing the NISQA Speech Naturalness~\cite{mittag2021nisqa}, 1,000 samples were randomly chosen as the test dataset. Additionally, we conduct a subjective audio evaluation using a 5-point Mean Opinion Score (MOS) test, involving 30 examples per model and 20 participants.
\newline
\textbf{Speech Enhancement} Our experiments were conducted using the VoiceBankDEMAND dataset~\cite{valentini2016investigating}. The dataset, derived from the VoiceBank corpus~\cite{veaux2013voice}, encompasses 30 speakers and is bifurcated into a training set with 28 speakers and a testing set with 2 speakers.The training utterances are deliberately mixed with eight real-recorded noise samples from the DEMAND database, in addition to two synthetically generated noise samples, at SNR levels of 0, 5, 10, and 15 dB. This results in a total of 11,572 training utterances.

For testing, the utterances are combined with different noise samples at SNR levels of 2.5, 7.5, 12.5, and 17.5 dB, culminating in a total of 824 testing utterances. Our algorithm was evaluated using the Perceptual Evaluation of Speech Quality (PESQ) and a deep learning evaluation approach, DNSMos~\cite{dubey2023icassp}.

\begin{algorithm}[t]
\caption{Wavelet CDiffuSE Training}
\begin{algorithmic}[1]
\For{$i = 1, 2, \dots, N_{\text{iter}}$}
    \State Sample $(x_0, x_n) \sim q_{\text{data}}, \epsilon \sim \mathcal{N}(0, \mathbf{I}),$
    \State $y_0 = DWT(x_0), y_n = DWT(x_n)$
    \State $t \sim \text{Uniform}(\{1, \dots, T\})$
    \State $y_t = ((1 - m_t) \sqrt{\bar{\alpha}_t} y_0 + m_t \sqrt{\bar{\alpha}_t} y_n) + \sqrt{\delta_t} \epsilon$
    \State Take gradient step on $\nabla_{\theta} \Big\| \frac{1}{\sqrt{1-\bar{\alpha}_t}} (m_t \sqrt{\bar{\alpha}_t} (y_n - y_0) + \sqrt{\delta_t} \epsilon) - \epsilon_{\theta}(y_t, y_n, t) \Big\|_2^2$
\EndFor
\end{algorithmic}
\label{Wavelet CDiffuSE Training}
\vspace{-0.2cm}
\end{algorithm}
\vspace{-0.1cm}
\subsection{Model Architecture and Training}
\vspace{-0.1cm}
To ensure a fair comparison with the baseline, we adhered to the identical parameter settings utilized in both Diffwave and CDiffuSE. To more effectively validate the versatility of our method, we conducted tests on both the base and large versions of Diffwave and CDiffuSE. To explore the distinct characteristics of various wavelets, we conducted experiments using a computational base of 32 NVIDIA V100 32GB GPUs. we conducted tests with different wavelets base using 32 V100 32G, including Haar, Biorthogonal 1.1 (bior1.1), Biorthogonal 1.3 (bior1.3), Coiflets 1 (coif1)~\cite{daubechies1988orthonormal}, Daubechies 2 (db2), and Cohen-Daubechies-Feauveau 5/3 (cdf53)~\cite{sullivan2003general}. The details of the parameter setting can be seen in Appendix~\ref{Experiment Setup}.
\subsection{Main Result}
\vspace{-0.1cm}
Table~\ref{Main Result} shows the results for various wavelet bases in both Speech Enhancement and Speech Synthesis tasks. It can be observed that, across all tasks, regardless of the type of wavelet basis used,  the training time, the inference time, and the required GPU memory consumption have been reduced by nearly half. In the Speech Enhancement task, when evaluated using the pseq metric, most wavelets, with the exception of the Coif1, performed comparably to the original model. The \textbf{DB2 wavelet} exhibited the best performance on both the base and large models.

Despite nearly doubling in training and inference speeds, its performance was only marginally lower than the original model, with a difference of 0.051 and 0.021, respectively. However, when we switch to using the DNSMos metric for evaluation, the scenario changes completely. When evaluating with the DNSMos metric, there is a complete shift in results. The \textbf{Coif1 wavelet} becomes the best performer. In the base model, it surpasses the original model by 0.009, and in the large model, the lead extends to 0.056. A detailed analysis will be presented in the subsequent sections.

In the task of Speech Synthesis, the results show some variations. In the base model, the Coif1 wavelet still outperforms others, even exceeding the original model by 0.004 in Speech Naturalness (SN). However, when we examine the large model, we find that although the Coif1 wavelet continues to perform well, it is the Bior1.3 wavelet that stands out as the top performer, surpassing the original model by 0.008 in terms of SN.

Through these experiments, we have demonstrated that our method can double the training and inference speeds of the speech diffusion model while achieving results that are comparable to, or even surpass, those of the original model. The consistent performance across both base and large models further validates the generalizability of our approach. The stable results on Diffwave and CDiffuSE highlight the versatility of our method across various tasks. This advancement enables the practical application of diffusion models in the field of speech, especially the accelerated training aspect, making it feasible to customize voices and perform targeted noise reduction for specific scenarios.
\vspace{-0.2cm}
\section{Further Study}
\vspace{-0.2cm}
Under the significant acceleration achieved by our method, we explore the potential for enhancing the quality of samples through wavelet transformation and further accelerating the training and sampling process of the diffusion model.
\vspace{-0.2cm}
\begin{table*}[t]
\centering
\setlength{\tabcolsep}{2pt} 
\renewcommand{\arraystretch}{0.5}
\begin{tabular}{@{}ccccccccc@{}}
\toprule
\multicolumn{5}{c|}{\textbf{Speech Enhancement}}                                              & \multicolumn{4}{c}{\textbf{Speech Synthesis}} \\ \midrule
\multicolumn{9}{c}{\cellcolor[HTML]{D9D8D8}\textbf{Base}}                                                                                    \\ \midrule
\multicolumn{1}{c|}{} &
  \textbf{PESQ $\uparrow$} &
  \textbf{DNS\_MOS $\uparrow$} &
  \textbf{Training Time$\downarrow$} &
  \multicolumn{1}{c|}{\textbf{RTF}$\downarrow$} &
  \textbf{MOS $\uparrow$} &
  \textbf{SN $\uparrow$} &
  \textbf{Training Time$\downarrow$} & 
  \textbf{RTF$\downarrow$} \\ \midrule
\multicolumn{1}{c|}{\textbf{Orignial}} & 2.466 & 3.116 & 481.784 & \multicolumn{1}{c|}{0.728} & 4.38±0.08    & 4.372    & 330.857    & 0.599   \\ \midrule \midrule
\multicolumn{1}{c|}{\textbf{Haar}}     & 2.387 & 3.008 & 248.065 & \multicolumn{1}{c|}{0.402} & 4.32±0.09    & 4.302    & 171.914    & 0.317   \\ \midrule
\multicolumn{1}{c|}{\textbf{Bior1}}    & 2.389 & 3.031 & 248.112 & \multicolumn{1}{c|}{0.402} & 4.33±0.06    & 4.300    & 172.077    & 0.317   \\ \midrule
\multicolumn{1}{c|}{\textbf{Coif1}}    & 1.625 & \textbf{3.125} & 248.997 & \multicolumn{1}{c|}{0.407} & \textbf{4.37±0.07}    & \textbf{4.376}    & 171.964    & 0.325   \\ \midrule
\multicolumn{1}{c|}{\textbf{DB2}}      & \textbf{2.415} & 3.032 & 251.215 & \multicolumn{1}{c|}{0.409} & 4.30±0.08    & 4.351    & 172.266    & 0.327   \\ \midrule
\multicolumn{1}{c|}{\textbf{Cdf53}}    & 2.367 & 3.049 & 249.190 & \multicolumn{1}{c|}{0.407} & 4.23±0.07    & 4.372    & 172.266    & 0.325   \\ \midrule
\multicolumn{1}{c|}{\textbf{Bior1.3}}  & 2.302     & 3.027     & 259.831       & \multicolumn{1}{c|}{0.413}     & 4.32±0.09   & 4.331         & 181.914     & 0.342       \\ \midrule
\multicolumn{9}{c}{\cellcolor[HTML]{D9D8D8}\textbf{Large}}                                                                                   \\ \midrule
\multicolumn{1}{c|}{\textbf{Original}} & 2.514 & 3.140 & 997.688 & \multicolumn{1}{c|}{6.387} & 4.41±0.08    & 4.395    & 806.158    & 6.055\\ \midrule \midrule
\multicolumn{1}{c|}{\textbf{Haar}}     & 2.463 & 3.127 & 507.813 & \multicolumn{1}{c|}{3.366} & \textbf{4.40±0.07}    & 4.229    & 408.123    & 3.061\\ \midrule
\multicolumn{1}{c|}{\textbf{Bior1}}    & 2.468 & 3.140 & 504.313 & \multicolumn{1}{c|}{3.363} & 4.33±0.07    & 4.360    & 408.132    & 3.060\\ \midrule
\multicolumn{1}{c|}{\textbf{Coif1}}    & 1.660 & \textbf{3.196} & 511.689 & \multicolumn{1}{c|}{3.443} & 4.39±0.06    & 4.351    & 412.727    & 3.152\\ \midrule
\multicolumn{1}{c|}{\textbf{DB2}}   & \textbf{2.493} & 3.125 & 513.384 & \multicolumn{1}{c|}{3.445} & 4.35±0.07    & 4.374    & 413.210    & 3.144\\ \midrule
\multicolumn{1}{c|}{\textbf{Cdf53}}    & 2.475 & 3.136 & 512.544 & \multicolumn{1}{c|}{3.440} & 4.31±0.06    & 4.325    & 412.963    & 3.149\\ \midrule
\multicolumn{1}{c|}{\textbf{Bior1.3}}  &2.395     & 3.126     & 519.353       & \multicolumn{1}{c|}{3.467}     & 4.32±0.09    & \textbf{4.403}       & 421.415      & 3.373\\ \midrule
\multicolumn{1}{c|}{\textbf{GT}}  &-- &-- &-- &\multicolumn{1}{c|}{--} &4.53±0.06  &--  &-- &-- \\ \bottomrule
\end{tabular}
\caption{The table presented above displays the results for various wavelet bases in both Speech Enhancement and Speech Synthesis tasks. SN represents Speech Naturalness. GT stands for Ground Truth, referring to the raw audio from human. 'Training Time' represents the time required for training in a single epoch(seconds). 'RTF' (Real-Time Factor) is utilized as a metric to assess inference time.}
\vspace{-0.3cm}
\label{Main Result}
\end{table*}
\begin{table}[t]
\centering
\begin{tabular}{@{}cccc@{}}
\toprule
\multicolumn{4}{c}{\textbf{Speech Synthesis (Haar Base)}} \\ \midrule
{Model} & \textbf{MOS} & \textbf{Training Time} & \textbf{RTF} \\
\midrule
GT       & 4.53±0.06 &-- &-- \\
Original & 4.38±0.08 & 330.857 & 0.599 \\
Haar2C     & 4.41±0.09 & 173.198 & 0.318 \\
Haar4C    & 4.32±0.09 & 65.350 & 0.126 \\
\bottomrule
\end{tabular}
\caption{The Table shows the result of  Multi-level wavelet Accelerator, the 4C means the speech signal will be decomposed into 4 Parts.}
\vspace{-0.5cm}
\label{Multilayer result}
\end{table}
\subsection{Low-frequency Speech Enhancer}
\begin{figure}[t]
    \centering
    \includegraphics[width=0.2\textwidth]{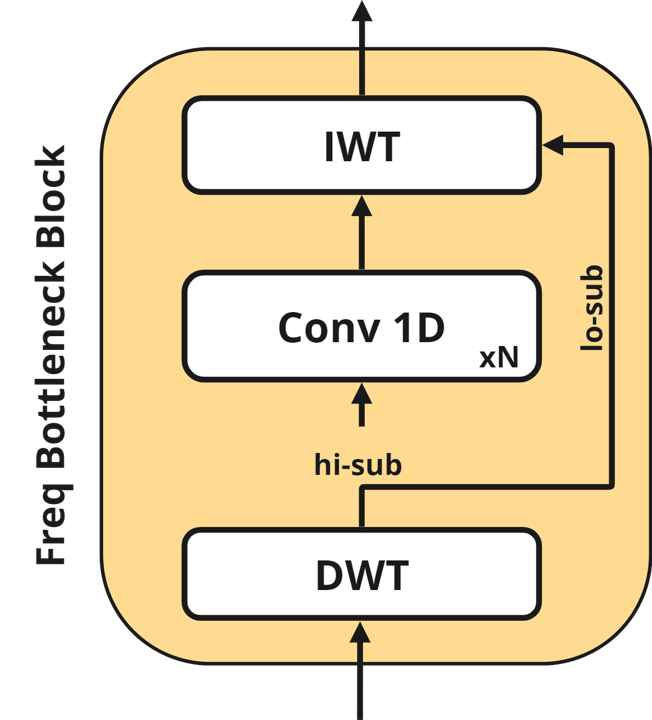}
    \caption{Overview of Frequency Bottleneck Block}\vspace{-5mm}
    \label{fig:fbb}
\end{figure}

In speech signals, the primary speech components are typically concentrated in the low-frequency range, while background noise tends to dominate the high-frequency spectrum~\cite{flanagan2013speech}. Therefore, to further enhance the quality of synthesized speech, we fully leverage the properties of wavelet decomposed signals. By performing Discrete Wavelet Transform (DWT) on the speech signals~\cite{shensa1992discrete}, we obtain a 2-channel vector, consisting of detail coefficients filtered through a high-pass filter and approximation coefficients filtered through a low-pass filter. Prior to feeding into the diffusion model, this vector is processed through the Frequency Bottleneck Block as shown in Figure~\ref{fig:fbb}, which amplifies the low-frequency speech signals and attenuates the background noise. Since different wavelet signals emphasize various speech characteristics during DWT, we tested six types of wavelets, as shown in Table~\ref{CNN Result}. The results indicate that the Haar wavelet, which focuses on signal discontinuities and rapid changes~\cite{stankovic2003haar}, achieves superior sampling quality compared to DiffWave after processing through the Frequency Bottleneck Block module. 
\vspace{-0.1cm}
\subsection{Multi-Level Wavelet Accelerator}
\begin{table*}[t]
\centering
\setlength{\tabcolsep}{2pt} 
\renewcommand{\arraystretch}{0.5}
\begin{tabular}{@{}ccccccccc@{}}
\toprule
\multicolumn{5}{c|}{\textbf{Speech Enhancement}}                                              & \multicolumn{4}{c}{\textbf{Speech Synthesis}} \\ \midrule
\multicolumn{9}{c}{\cellcolor[HTML]{D9D8D8}\textbf{Base}}                                                                                    \\ \midrule
\multicolumn{1}{c|}{} &
  \textbf{PESQ $\uparrow$} &
  \textbf{DNS\_MOS $\uparrow$} &
  \textbf{Training Time$\downarrow$} &
  \multicolumn{1}{c|}{\textbf{RTF}$\downarrow$} &
  \textbf{MOS $\uparrow$} &
  \textbf{SN $\uparrow$} &
  \textbf{Training Time$\downarrow$} & 
  \textbf{RTF$\downarrow$} \\ \midrule
\multicolumn{1}{c|}{\textbf{Orignial}} & 2.466 & 3.116 & 481.784 & \multicolumn{1}{c|}{0.728} & 4.38±0.08    & 4.372    & 330.857    & 0.599   \\ \midrule
\multicolumn{1}{c|}{\textbf{Haar}}     & \textbf{2.477} & \textbf{3.157} & 249.2735 & \multicolumn{1}{c|}{0.405} & \textbf{4.41±0.09}    & \textbf{4.421}   & 173.19   & 0.317   \\ \midrule
\multicolumn{1}{c|}{\textbf{Bior1}}    & 2.429 & 3.118 & 251.908 & \multicolumn{1}{c|}{0.405} & 4.36±0.08    & 4.353    & 171.490    & 0.318   \\ \midrule
\multicolumn{1}{c|}{\textbf{Coif1}}    & 1.647 & 3.129 & 250.579 & \multicolumn{1}{c|}{0.410} & 4.38±0.06    & 4.104    & 171.455    & 0.327   \\ \midrule
\multicolumn{1}{c|}{\textbf{DB2}}      & 2.463 & 2.999 & 251.004 & \multicolumn{1}{c|}{0.411} & 4.36±0.07    & 4.252   & 171.777    & 0.328   \\ \midrule
\multicolumn{1}{c|}{\textbf{Cdf53}}    & 2.412 & 3.027 & 251.686 & \multicolumn{1}{c|}{0.410} & 4.27±0.06    & 4.327    & 173.427    & 0.327   \\ \midrule
\multicolumn{1}{c|}{\textbf{Bior1.3}}  & 2.463    & 3.014     & 258.316       & \multicolumn{1}{c|}{0.421}     & 4.34±0.07   & 4.342         & 182.731     & 0.333       \\ \midrule
\multicolumn{9}{c}{\cellcolor[HTML]{D9D8D8}\textbf{Large}}                                                                                   \\ \midrule
\multicolumn{1}{c|}{\textbf{Original}} & 2.514 & 3.140 & 997.688 & \multicolumn{1}{c|}{6.387} & 4.41±0.08    & 4.395    & 806.158    & 6.055\\ \midrule
\multicolumn{1}{c|}{\textbf{Haar}}     & 2.463 & 3.127 & 507.813 & \multicolumn{1}{c|}{3.366} & 4.34±0.06    & 4.229    & 408.123    & 3.061\\ \midrule
\multicolumn{1}{c|}{\textbf{Bior1}}    & 2.468 & 3.140 & 504.313 & \multicolumn{1}{c|}{3.363} & 4.35±0.07    & 4.360    & 408.132    & 3.060\\ \midrule
\multicolumn{1}{c|}{\textbf{Coif1}}    & 1.660 & \textbf{3.196} & 511.689 & \multicolumn{1}{c|}{3.443} & 4.35±0.08    & 4.351    & 412.727    & 3.152\\ \midrule
\multicolumn{1}{c|}{\textbf{DB2}}   & \textbf{2.493} & 3.125 & 513.384 & \multicolumn{1}{c|}{3.445} & 4.37±0.07    & 4.374    & 413.210    & 3.144\\ \midrule
\multicolumn{1}{c|}{\textbf{Cdf53}}    & 2.475 & 3.136 & 512.544 & \multicolumn{1}{c|}{3.440} & \textbf{4.43±0.09}    & 4.325    & 412.963    & 3.149\\ \midrule
\multicolumn{1}{c|}{\textbf{Bior1.3}}  &2.395     & 3.126     & 522.733       & \multicolumn{1}{c|}{3.483}     & 4.38±0.06    & \textbf{4.403}       & 422.326      & 3.342\\ \bottomrule
\end{tabular}
\caption{The table presented above displays the results for various wavelet bases in both Speech Enhancement
and Speech Synthesis tasks. SN represents Speech Naturalness. ’Training Time’ represents the time required for training in a single epoch(seconds). ’RTF’
(Real-Time Factor) is utilized as a metric to assess inference time.}
\label{CNN Result}
\end{table*}
To further enhance training and sampling speeds, we implemented a multi-level DWT approach, as demonstrated in Figure~\ref{fig:dwt}. This method reduces the length of speech signal features to a quarter of their original size, and increases the channel count to four. Concurrently, the Frequency Bottleneck Block, designed to intensify speech signals, is expanded into the Multi-level Low-Frequency Voice Enhancement Module, which encompasses a multi-level residual block. This block is adept at progressively attenuating high-frequency components, as depicted in Figure~\ref{fig:mulcnn}. This methodology significantly reduces both training and sampling times, with training speeds approximately five times faster than the original DiffWave and sampling speeds about three times quicker. As shown in Table~\ref{Multilayer result}, the Mean Opinion Score (MOS) indicates that the audio quality of the samples remains comparably high, which underscores its strong practicality.
\vspace{-0.1cm}
\section{Ablation Study and Analysis}
\subsection{Effect of Vanishing Moments, Smoothing and Complexity}
From Table~\ref{Main Result}, it can be observed that Coif1 performs well on the DNSmos metric and in speech synthesis tasks, yet exhibits poor performance when evaluated using the PSEQ. The difference between DNSmos and PSEQ lies in the fact that DNSmos does not require reference audio; it is used directly to evaluate the quality of the generated speech. After listening to several sets of generated speech, we discovered that while the diffusion model using Coif1 wavelets produces clear and smooth speech, there is a significant alteration in timbre compared to the original sound. By comparing with DB2 and Haar wavelets, we can conclude that as the vanishing moment increases and complexity follows (Coif1 > DB2 > Haar), the diffusion model tends to generate clearer and smoother speech. However, once the vanishing moment reaches a certain level, the timbre of the sound is altered. This characteristic enables the selection of Coif1 wavelets in scenarios where only noise reduction is needed, or in speech synthesis tasks where timbre is of lesser concern and the emphasis is on naturalness.
\vspace{-0.1cm}
\subsection{Effect of Order of the Wavelet}
Comparing bior1.1 with bior1.3, we observe that with an increase in the reconstruction order, both the PSEQ and DNS\_MOS scores decrease. This indicates that as the reconstruction order rises, the diffusion model's ability to handle noise diminishes, although there is a slight improvement in speech synthesis tasks. We believe this is because bior1.3, compared to bior1.1, captures more high-frequency information. However, noise compared to human voice generally occupies the high-frequency range, which explains why bior1.3 performs less effectively than bior1.1 in speech enhancement tasks.

Comparing Haar (DB1) with DB2, we find that when the reconstruction order remains the same, an increase in the decomposition order enhances the performance of the wavelet speech diffusion model, especially in terms of stability and superior performance in speech enhancement. It effectively removes noise while maintaining the timbre without significant changes.
In speech synthesis tasks, DB2 also shows improvement over Haar, which we attribute to the increased complexity of the wavelet.
\vspace{-0.5cm}
\subsection{Relationship between Wavelet base and Training/Inference Speed}
From Table~\ref{Main Result}, it is evident that regardless of the wavelet used, both training and inference speeds are nearly doubled compared to the original model. The table indicates that when wavelets are applied to the diffusion model, Haar and bior1.1 exhibit similar speeds. The differences in speed between Coif1, DB2, and cdf53 are minimal, with bior1.3 being the slowest. We discovered that their speeds do not strictly correlate with their computational complexity. Our analysis suggests that the longer filter length of Bior1.3 in implementation, combined with the inherently long nature of speech signals, results in increased computational overhead.

\subsection{Effect of Frequency Enhancer}
After incorporating the Frequency Enhancer, most wavelet speech diffusion models showed an improvement in performance. In speech enhancement tasks, Haar, bior1.3, and cdf53 wavelets demonstrated significant improvements. Meanwhile, the training and inference speeds, compared to the wavelet diffusion model without the Frequency Enhancer, remained virtually unchanged, falling within the margin of error. Haar and Coif1 wavelets diffusion model even outperformed the original model, indicating that by simply adding a small pre-processing module, we can surpass the performance of the original model while significantly increasing training and inference speeds. However, we believe that the reasons for the performance enhancement offered by these three wavelets are not the same. 
For the Haar wavelet, its ability to capture discontinuities and abrupt changes in signals makes it particularly effective at handling non-stationary signals like speech. The Frequency Enhancer further amplifies this capability. Bior1.3, due to its enhanced ability to capture high-frequency signals, sees a reduction in noise after processing with the Frequency Enhancer. Therefore, its performance improves compared to when the Frequency Enhancer is not used. For the cdf53 wavelet, it is capable of compressing signals with minimal loss. After being enhanced by the Frequency Enhancer, high-frequency noise is effectively removed, while low-frequency signals are well preserved. This lossless property is better demonstrated in the field of speech synthesis, where, after enhancement by the Frequency Enhancer, the performance slightly exceeds that of the original model in MOS tests. For detailed data, please refer to table~\ref{CNN Result}.
\vspace{-0.1cm}
\subsection{Effect of Multi-Level Wavelet Accelerator}
To further explore the potential for acceleration, we conducted tests in the field of speech synthesis using the Haar wavelet, which demonstrated the most stable performance. The results of the experiment are shown in Table~\ref{Multilayer result}. It can be observed that when the speech signal is split into quarters of its original length, both training and inference speeds increase by more than fivefold. However, unlike the results of splitting just once (as shown in the second row of Table~\ref{Multilayer result}, corresponding to the second row of Table~\ref{CNN Result}), which were better than the original model, the results after splitting four times, even with the Frequency Enhancer, exhibited a notable decline in MOS values. We believe this is due to information loss caused by excessive compression. However, the substantial increase in speed still makes this method worth considering for scenarios where ultra-clear audio is not required.

\subsection{Performance on Multi-Speaker Dataset}
\begin{table}[t]
\centering
\setlength{\tabcolsep}{2pt} 
\renewcommand{\arraystretch}{0.8} 
\scalebox{0.9}{ 
\begin{tabular}{@{}lccc@{}}
\toprule
\textbf{Model on VCTK dataset} & \textbf{PESQ} & \textbf{SN} & \textbf{RTF} \\ \midrule
ori base  & 4.2179  & 3.1165  & 0.9072 \\
haar base & 4.2069  & 3.1209  & 0.3957 \\
bior1.1 base & 4.0828  & 3.1473  & 0.4077 \\
bior1.3 base & 4.0658  & 3.1059  & 0.3987 \\
coif1 base & 4.2025  & 2.9393  & 0.4031 \\
cdf53 base & 4.1089  & 3.1937  & 0.3843 \\
db2 base & 4.1634  & 2.9744  & 0.4034 \\ \midrule
haar base* & 4.2323  & 3.0138  & 0.4147 \\
bior1.1 base* & 4.2083  & 3.0415  & 0.3943 \\
bior1.3 base* & 4.1921  & 3.0551  & 0.3995 \\
coif1 base* & 4.1824  & 3.0406  & 0.4034 \\
cdf53 base* & 4.0939  & 3.2039  & 0.3949 \\
db2 base* & 4.1601  & 3.0479  & 0.4053 \\ \bottomrule
\end{tabular}
}
\caption{Low-frequency Speech Enhancer results on VCTK dataset. RTF (Real-Time Factor) is utilized as a metric to assess inference time. SN denotes Speech Naturalness, * denotes results from Low-frequency Speech Enhancer}
\vspace{-12pt}
\label{VCTK Results}
\end{table}
In response to concerns regarding the generalizability of our method, we conducted additional experiments using the VCTK dataset~\cite{oord2016wavenet}, applying all the wavelets tested in our original study. To further strengthen our findings, we also evaluated the performance of our low-frequency speech enhancer, which forms part of our ongoing research efforts, on the same dataset. The results, presented in Table \ref{VCTK Results}, demonstrate that our approach maintains consistent performance across different datasets.

\section{Conclusion}

In this paper, we have enhanced the speech diffusion model by transitioning its generation target to the wavelet domain, thereby doubling the model's training and inference speeds. We offer a new perspective on accelerating speech models by focusing on processing the signal itself rather than modifying the model. Our approach has demonstrated model versatility and task adaptability across both speech enhancement and synthesis. Through our research, we found that the Coif1 wavelet is an excellent choice for scenarios requiring noise reduction without the need to preserve timbre, while the DB2 wavelet is preferable when changes in timbre must be considered. For speech synthesis tasks, the Haar wavelet offers simplicity and effectiveness, whereas the cdf53 wavelet excels at preserving information to the greatest extent. Additionally, We designed two simple and easily integrable front-end modules. The first achieves better performance than the original model while doubling the speed. The second offers a performance comparable to the original while enabling an acceleration of more than five times.

\section*{limitations}
In this study, speed tests were conducted on a large-scale cluster, subject to the hardware variability inherent in the cluster (despite all GPUs being V100s, they may not be identical), which could introduce some timing inaccuracies. However, considering that the training and inference times for most wavelet-utilizing diffusion models do not significantly differ, we believe these discrepancies can be disregarded. This does not detract from our contribution of accelerating the speech diffusion model by a factor of two.

\section*{Ethics Statement}
Our proposed model diminishes the necessity for high-quality speech synthesis, potentially affecting employment opportunities for individuals in related sectors, such as broadcasters and radio hosts. By lowering the training costs, our approach may impact a broader audience.

\bibliography{custom}
\bibliographystyle{acl_natbib}

\appendix

\section{Details of Experiment Setup}
\label{Experiment Setup}
Diffwave offers two configurations: base and large. In the base version, the model comprises 30 residual layers, a kernel size of 3, and a dilation cycle of [1, 2, ..., 512]. It utilizes 50 diffusion steps and a residual channel count of 64. The large version maintains all parameters identical to the base, except for an increase to 128 residual channels and 200 diffusion steps. All models employed the Adam optimizer, with a batch size of 16 and a learning rate of $2 \times 10^{-4}$.We trained each DiffWave model for a total of 1 million steps.

We conducted evaluations on two versions of CDiffuSE: base and large. The base CDiffuSE model employs 50 diffusion steps, while the large CDiffuSE model uses 200 diffusion steps. Batch sizes differ, with the base CDiffuSE set to 16 and the large CDiffuSE set to 15. Both the base and large CDiffuSE models were trained for 300,000 iterations, following an early stopping scheme.

\section{Details of CDiffuSE}
\label{Appendix CDiffuSE}
The CDiffuSE is trying to optimize the likelihood by ELBO condition for the conditional diffusion process. we further extend it to the Wavelet Latent domain.
\begin{equation}
\scalebox{0.7}{$
\begin{aligned}
ELBO = & -\mathbb{E}_q \left( D_{KL}(q_{\text{cdiff}}(\mathbf{y}_T|\mathbf{y}_0, y_n) \parallel p_{\text{latent}}(\mathbf{y}_T|y_n)) \right) \\
& + \sum_{t=2}^{T} D_{KL}(q_{\text{diff}}(\mathbf{y}_{t-1}|\mathbf{y}_t, \mathbf{y}_0, y_n) \parallel p_{\theta}(\mathbf{y}_{t-1}|\mathbf{y}_t, y_n)) \\
& - \log p_{\theta}(\mathbf{y}_0|\mathbf{y}_1, y_n).
\end{aligned}
$}
\end{equation}

Parameters $c_{y_t}$, $c_{y_n}$, and $c_{\epsilon_t}$ be derived as:

\begin{equation}
\scalebox{0.8}{$
\begin{aligned}
c_{yt} &= \frac{1 - m_t}{1 - m_{t-1}} \frac{\delta_{t-1}}{\delta_t} \sqrt{\alpha_t} + (1 - m_{t-1}) \frac{\delta_{t|t-1}}{\delta_t} \frac{1}{\sqrt{\alpha_t}}, \\
c_{yn} &= \frac{(m_{t-1}\delta_t - m_t(1 - m_t)\alpha_t\delta_{t-1})\sqrt{\hat{\alpha}_{t-1}}}{1 - m_{t-1}\delta_t}, \\
c_{\epsilon_t} &= \frac{(1 - m_{t-1})}{\delta_t} \frac{\delta_{t|t-1}\sqrt{1 - \hat{\alpha}_t}}{\sqrt{\alpha_t}}.
\end{aligned}
$}
\end{equation}

Where $\delta_{t}$ variance term, all other parameters have been mentioned in main section.

\section{Details of Wavelet Diffusion Accelerator}
\label{Wavelet Diffusion Accelerator}

\subsection{How Wavelets Accelerate Diffusion models}
In \S\ref{WaveletCompression}, we detailed the application of Discrete Wavelet Transform (DWT) and Inverse Discrete Wavelet Transform (IWT) in processing audio signals, highlighting how these techniques compress the audio signal features during the diffusion process. This section elaborates on the principles behind the acceleration offered by the Wavelet Diffusion Accelerator. 

To facilitate training acceleration, the diffusion model shifts its focus from generating complete audio signals with extensive features to producing compressed speech signals in wavelet domain. In line with this shift, DWT is employed to process the raw audio signal $g\left( n \right) \in \mathbb{R}^{1\times 2x}$, where $n$ denotes the sample index, through two complementary filters. Specifically, a low-pass filter $\phi$ extracts the low-frequency components $\varPsi_{low}\in \mathbb{R}^{1\times 2x}$:
\begin{equation}
    \varPsi_{low}\left( n \right) =\sum_{k=-\infty}^{+\infty}{g\left( k \right) \phi \left( 2n-k \right)}.
\end{equation}
And a high-pass filter $\psi$ is utilized to extract the high-frequency portion $\varPsi_{high}\in \mathbb{R}^{1\times 2x}$:
\begin{equation}
    \varPsi _{high}\left( n \right) =\sum_{k=-\infty}^{+\infty}{g\left( k \right) \psi \left( 2n-k \right)}.
\end{equation}
To further reduce the size of the features and emphasize the signal's essential characteristics, downsampling is applied to both parts of the signal, resulting in the approximation coefficients $cA$ and the detail coefficients $cD$:
\begin{equation}
    cA=\varPsi _{low}\downarrow 2,
\end{equation}
\begin{equation}
    cD=\varPsi _{high}\downarrow 2.
\end{equation}
At this stage, the signal $g\left( n \right) \in \mathbb{R}^{1\times 2x}$ is compressed into $h\left( n \right) \in \mathbb{R}^{2\times x}$, wherein $h$ embodies a two-channel structure, each channel containing features of halved length.

This change significantly contributes to reducing the computational time required for training the diffusion model. To further demonstrate, we exemplify with the computational changes in the diffusion model's first convolutional layer. Assuming the output channel count is 
$C_{out}$, the kernel size is $K$, and the output length $L_{out}$ remains unchanged from the input length. The formula for calculating Multiply-Accumulate Operations (MACs) per channel is:
\begin{equation}
    MAC_{each}=K\times C_{out}\times L_{out}.
\end{equation}
Hence, for each channel, with $h(n)$ as the input, the computational load in the first convolutional layer is halved:
\begin{equation}
    MAC_{h\left( n \right)}=K\times C_{out}\times x
    \\
    =\frac{1}{2}MAC_{g\left( n \right)}.
\end{equation}
Given the GPU's optimization for parallel computing, the increase in the number of channels does not lead to a linear increase in computational time. From experimental results, both training and sampling times of the diffusion model have a significant reduction.

\subsection{Wavelets for Diffusion Acceleration: Why Not FFT}
While wavelet and Fourier transforms both serve as essential tools in signal processing and share similarities in handling time and frequency domain information, this section explores why Fast Fourier Transform (FFT) is not applicable for accelerating diffusion models. This is determined by the inherent nature of the Fourier transform. Assuming $f(t)$ is the representation of the signal in the time domain and $\hat{f}(\omega)$  is its representation in the frequency domain, where $t$ stands for time and $\omega$ for frequency, then the CFT can be described as:
\begin{equation}
    \hat{f}\left( \omega \right) =\int_{-\infty}^{+\infty}{f\left( t \right) e^{-i\omega t}dt}.
\end{equation}
The Fourier transform fits the entire signal $f(t)$ with a series of sine and cosine functions, converting it into frequency domain information $\hat{f}\left( \omega \right)$. As a result, the signal is stripped of time information following this transformation. However, conventional input audio signals $f(t)$ display traits where local frequency domain features shift in response to variations in short-time segments of the time domain signal, like abrupt transitions or displacements. This lack of capability to concurrently analyze local time and frequency domain information makes the Fourier transform insufficient for accurately recreating the original audio in generative models.

In contrast, for the wavelet transform, assuming $\psi \left( t \right) $ as a basic wavelet function, let:
\begin{equation}
    \psi _{a,b}\left( t \right) =\frac{1}{\sqrt{\left| a \right|}}\psi \left( \frac{t-b}{a} \right). 
\end{equation}
where $a,b\in \mathbb{R}$, $a\ne 0$, and the function $\psi _{a,b}\left( t \right)$ is called a continuous wavelet, generated from the mother wavelet $\psi \left( t \right) $ and dependent on parameters $a$ and $b$. Therefore, the continuous wavelet transform can be written as:
\begin{equation}
\hat{f}\left( a,b \right) =\frac{1}{\sqrt{\left| a \right|}}\int_{-\infty}^{+\infty}{f\left( t \right) \overline{\psi \left( \frac{t-b}{a} \right) }dt}.
\end{equation}
At this juncture, the wavelet transform converts a univariate time-domain signal $f(t)$ into a bivariate function $\hat{f}\left( a,b \right)$ encompassing both time and frequency domain information. It enables targeted analysis of local frequency domain characteristics corresponding to specific time domain segments, making it particularly well-suited for handling common non-stationary audio signals.

Besides, the wavelet transform's capability for time-frequency localization analysis ensures that downsampling and compressing $cA$ and $cD$ does not result in significant information loss. On the contrary, based on the Discrete Fourier Transform, FFT struggles with signal compression for diffusion acceleration due to its local frequency domain transformations affecting characteristics across the entire time domain.
\end{document}